\newcommand{\gtsim}{\mbox{{\raisebox{-0.4ex}{$\stackrel{>}{{\scriptstyle\sim}}
$}}}}

\documentclass[cup5b]{caps}
\usepackage{graphicx}
\usepackage{amssymb}
\usepackage{ociwsymp1e}
\HeadText{M. J. Jarvis and R. J. McLure}

\begin{document}

\pagenumbering{arabic}

\author[]{M. J. JARVIS$^1$ and R. J. MCLURE$^2$ \\
(1) Sterrewacht Leiden, Postbus 9513, 2300 RA Leiden, The Netherlands \\
(2) Institute for Astronomy, Edinburgh University, Edinburgh EH9 3HJ }

\chapter{The $M_{\rm BH} - L_{\rm rad}$ Relation for \\ Flat-Spectrum Quasars}

\begin{abstract}

In this proceedings we summarize our recent letter (Jarvis
\& McLure 2002) where we suggested that by correcting for the
inevitable effects of inclination, the black-hole masses of
flat-spectrum quasars (FSQ) with intrinsically powerful radio jets are confined,
virtually exclusively, to $M_{\rm BH} >10^{8}$~M$_{\odot}$. 
After considering realistic Doppler boosting factors, many of the FSQ would be more accurately classified as radio-intermediate or radio-quiet quasars.
This range in radio luminosity suggests that the FSQ are fully consistent with an upper boundary on radio power of
the form $L_{\rm 5GHz} \propto M_{\rm BH}^{2.5}$.
\end{abstract}

\section{Introduction}

One question which has recently received a great deal of
attention in the literature is whether or not the mass of an AGN's black
hole is strongly related to it's radio luminosity. This question is of
importance, because if it is established that radio-loud and
radio-quiet quasars have different black-hole mass distributions, it
may help explain why quasars of comparable optical luminosities can
differ in their radio luminosity by many orders of magnitude. On the
contrary, if radio-loud and radio-quiet quasars are found to have
essentially identical black-hole mass distributions, then the search
for the origin of radio loudness must move to some other physical
parameter such as black-hole spin.

Using the spectral data of Boroson \& Green (1992), Laor (2000)
investigated the relation between black-hole mass and radio
luminosity in the Palomar-Green quasar sample using the virial
black-hole mass estimator. The results from this 
analysis pointed to an apparent bi-modality in black-hole mass, 
with virtually all of the radio-loud quasars containing 
black holes with masses $M_{\rm BH} > 
10^{9}~$M$_{\odot}$, whereas the majority of quasars with black hole masses $M_{\rm BH} < 3 \times 10^{8}~$M$_{\odot}$ were radio quiet.

A similar result was arrived at by McLure \& Dunlop (2002) using a
sample of radio-loud and radio-quiet quasars matched in terms of both
redshift and optical luminosity. However, the
substantial over-lap between the black-hole mass distributions of the
two quasar samples indicated in addition that black-hole mass could
not be the sole parameter controlling radio power.

In contrast to the studies
outlined above, Ho (2002) suggested that there was
no clear relationship between radio power and black-hole mass, leading
the author to conclude that radio-loud AGN could be powered by black holes with a large
range of masses ($10^{6}\rightarrow
10^{9}{\rm M}_{\odot}$).

Following their study of the black-hole masses and host-galaxy
properties of low redshift radio-loud and radio-quiet quasars, Dunlop
et al. (2003) proposed an alternative view of the $M_{\rm BH}-L_{\rm rad}$ plane. 
They argue that the location of both active and
non-active galaxies on the $M_{\rm BH}-L_{\rm rad}$ plane appears to be 
consistent with the existence of an 
upper and lower envelope, both of the approximate form $L_{\rm 5GHz}
\propto M_{\rm BH}^{2.5}$, but separated by some 5 orders of magnitude in
radio power. In this scheme the upper and lower envelopes delineate 
the maximum and minimum radio luminosity capable of being produced by 
a black hole of a given mass. 

However, the recent study by Oshlack, Webster \& Whiting 
(2002; hereafter OWW02) of the black-hole masses of a sample
of flat-spectrum radio-loud quasars from the Parkes Half-Jansky Flat
Spectrum sample of Drinkwater et al. (1997), casts doubt on the
existence of any upper threshold in the $M_{\rm BH}-L_{\rm rad}$ plane. OWW02
found that their flat-spectrum quasars harbour black-hole masses in the range $10^{6}\rightarrow
10^{9}M_{\odot}$, and therefore lie well above the upper $L_{\rm 5GHz}
\propto M_{\rm BH}^{2.5}$ boundary proposed by Dunlop et al. (2003). The conclusion
reached by OWW02 following this result was that previous studies have
actively selected against including powerful radio sources with
relatively low black-hole masses, due to their concentration on luminous,
 optically selected radio-loud quasars.

Here we use the OWW02 sample to re-examine the position of
these flat-spectrum radio-loud objects on the $M_{\rm BH}-L_{\rm rad}$ plane 
when both Doppler boosting effects and 
the likely geometry of the broad-line region are taken into
account. 

\section{Measuring Black Hole Masses}\label{sec:virial}

The virial black-hole mass 
estimate uses the width of the broad Hydrogen Balmer emission 
lines to estimate the broad line region (BLR) velocity dispersion. The 
black hole mass may then be calculated under the assumption that 
the velocity of the BLR clouds is Keplerian:
\begin{equation}
M_{\rm BH} = R_{\rm BLR} V^{2} G^{-1},
\end{equation}
where $V$ is the velocity dispersion of the BLR clouds, usually 
estimated from the full-width half maximum (FWHM) of the H$\beta$ line, 
and $R_{\rm BLR}$ is the radius of the broad-line region.

The measurement of the radius of the BLR is ideally achieved by 
reverberation mapping, in which the continuum and line variations of a
number of sources are monitored over a number of years (e.g. Wandel, Peterson \& Malkan 1999; Kaspi et al. 2000).

Unfortunately, reverberation mapping of quasars is extremely time
consuming and it remains unrealistic that the black-hole masses of a
large sample of quasars can be measured in this way. However, the
radius of the BLR is found to be correlated with the monochromatic AGN
continuum luminosity at 5100\AA, $\lambda L_{5100}$ 
(e.g. Kaspi et al. 2000). Therefore, this correlation can be 
exploited to produce a virial black-hole mass estimate from a 
single spectrum covering the H$\beta$ emission line.
 
Here we adopt the calibration of the correlation between 
$R_{\rm BLR}$ and $\lambda L_{5100}$ from McLure \& Jarvis (2002; see also these proceedings), i.e.
\begin{equation}
R_{\rm BLR} = (25.4 \pm 4.4)[\lambda L_{5100} / 10^{37}{\rm W}]^{(0.61
\pm 0.10)},
\end{equation}
which, when combined with BLR velocity estimate from the $H\beta$
FWHM, leads to a black-hole mass estimate given by :
\begin{equation}
\frac{M_{\rm BH}}{\rm M_{\odot}} = 4.74 \left( \frac{\lambda L_{5100}}{10^{37} {\rm W}} \right )^{0.61} \left( \frac{{\rm FWHM}(H\beta)}{\rm km~s^{-1}}\right)^{2}.
\end{equation}
\noindent

\section{Doppler Boosting of the Radio Flux in Flat-Spectrum Radio Sources}\label{sec:doppler}

  \begin{figure}
    \centering
    \vspace{6cm}
\includegraphics{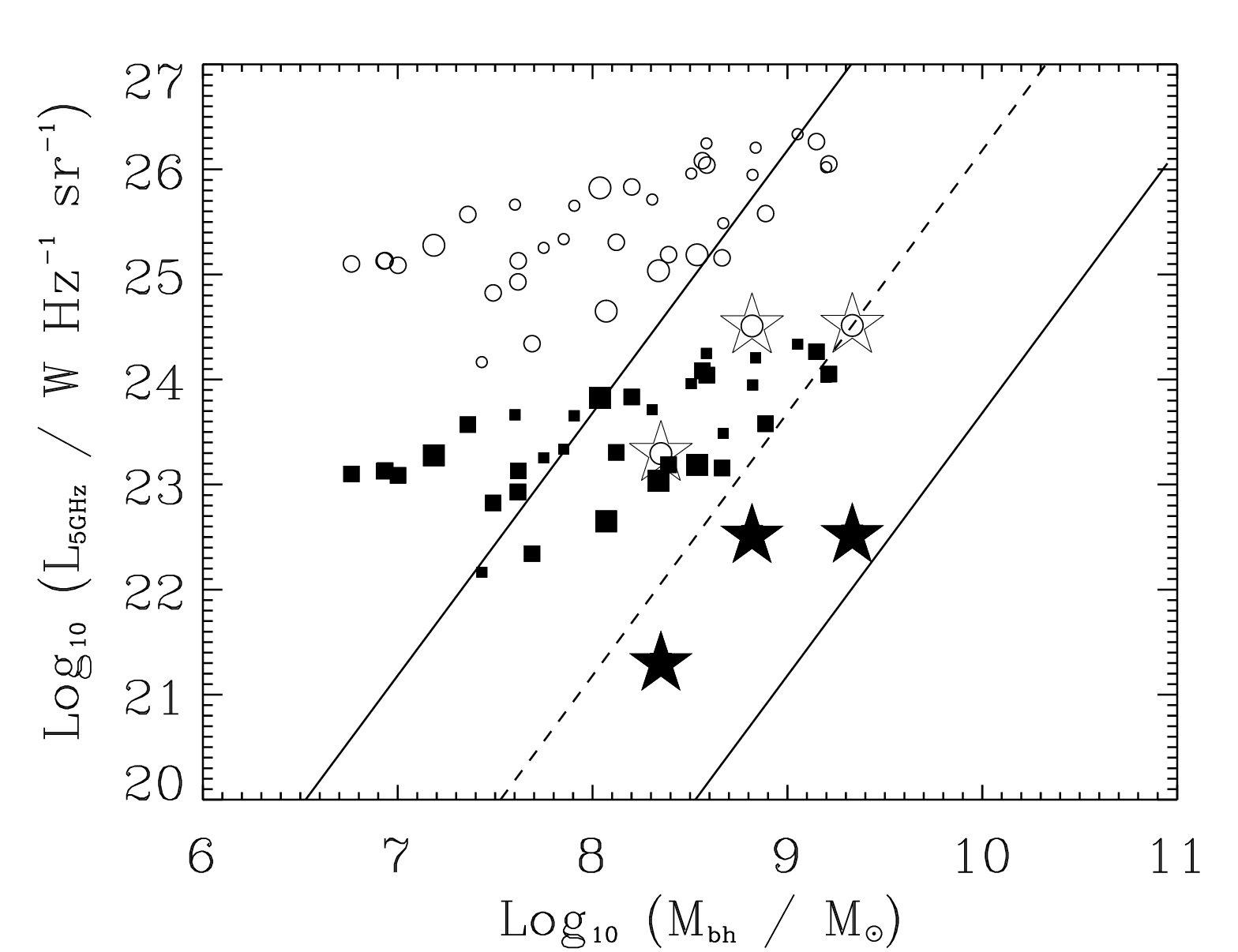}
\includegraphics{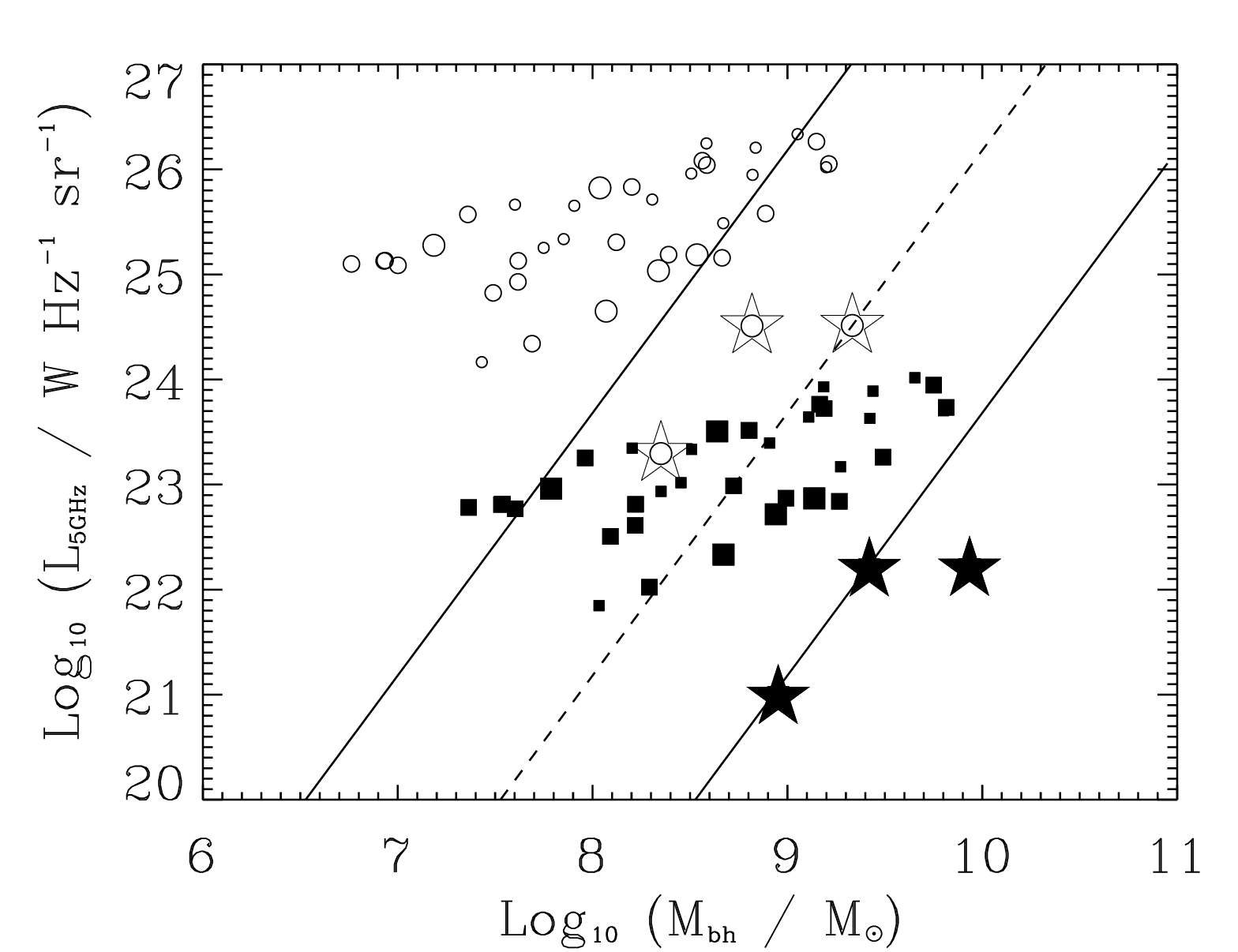}
    \caption{({\it left}) Total radio luminosity at 5~GHz versus
    black-hole mass for the FSQ from OWW02. Open circles are the
    original points of OWW02. Filled squares are the same sources with
    their radio flux density decreased by a factor of $\sim 100$, in
    accordance with the expected Doppler boosting factor. The size of
    the symbols are scaled according to radio spectral index and the
    stars are the three FSQ with the steepest spectral indices (see
    Jarvis \& McLure 2002 for a full description). The lines are
    relations of the form $L_{5 GHz} \propto M_{\rm BH}^{2.5}$ offset by
    2.5 dex from each other and represent the envelopes discussed in
    Dunlop et al. (2003). ({\it right}) The same plot but with a
    horizontal shift to account for the probable effect of orientation
    on the FWHM of the H$\beta$ broad emission line if the BLR has a
    disk-like geometry. }
    \label{fig1}
  \end{figure}

Flat-spectrum radio samples unavoidably contain a mix of radio source
populations including starbursts, Giga-Hertz Peaked Spectrum sources and Compact Steep Spectrum sources. However one population
is thought to dominate, the Doppler boosted sources. These are
preferentially selected in high-frequency samples because the
superposition of many synchrotron self-absorbed spectra along our
line-of-sight results in a flat-spectrum at high-radio frequencies.
As the radio emission is propagating along our line-of-sight in these
objects, the relativistic velocities associated with powerful radio
sources means that face-on radio
sources may undergo relativistic beaming which we see as a boost in
the flux.

From a statistical study of low-frequency and high-frequency selected
radio sources, Jackson \& Wall (1999) have shown that high-frequency
selected flat-spectrum sources have an opening angle within
$7^{\circ}$ of our line-of-sight. We note however that this value is
essentially a mean value and that both smaller or larger opening
angles are undoubtedly consistent with a Doppler boosting
paradigm. The opening angles may also depend on the intrinsic radio
power of the source (e.g. Jackson \& Wall 1999), thus the level of
Doppler boosting in a sample of flat-spectrum sources may have a wide
distribution.

However, keeping these caveats in mind, we can estimate the amount
of Doppler boosting the average flat-spectrum sources will exhibit,
compared with the Doppler boosting of the average quasar, if the
maximum opening angle is known for each population. Following the
method of Jarvis \& Rawlings (2000), we take the maximum opening angle
for which we observe a radio source as a quasar to be $53^{\circ}$, as
derived from the quasar fraction in low-frequency selected samples
(Willott et al. 2000). Consequently, averaging over solid angle, the
mean opening angle of steep spectrum radio-loud quasars is $\sim
37^{\circ}$ .

The boosting of the radio flux increases as
$\Gamma^{2}$:

\begin{equation}
\Gamma = \gamma^{-1} (1-\beta \cos \theta)^{-1} ,
\end{equation}

\noindent
where $\gamma$ is the Lorentz factor, $\beta=v/c$ and $\theta$ is the
angle between the radio jet and the line of sight. 
Therefore, adopting the conservative approach of 
substituting $\theta_{\rm flat} = 7^{\circ}$ for the flat-spectrum 
sources (many of the sources will have $\theta < 7^{\circ}$) and 
$\theta_{\rm steep} = 37^{\circ}$ for the steep spectrum quasar
population, we find that the Doppler
boosting factor is $\Gamma_{\rm flat}^{2}/\Gamma_{\rm steep}^{2}
\gtsim 100$. Hence, the intrinsic radio luminosity of the flat-spectrum
population is of the order $\gtsim 100$-times fainter than the
intrinsic radio luminosity of the average steep-spectrum radio-loud quasar.

In Fig.~\ref{fig1} we plot radio luminosity versus
black-hole mass with both the original radio luminosities, without any
boosting correction, and the same objects with the radio luminosity
decreased by a factor of 100. 
It is clear from Fig.~\ref{fig1} that following the correction for
Doppler boosting the vast majority of the flat-spectrum sources now
lie within the $L_{\rm rad} - M_{\rm BH}$ envelope of the quasar
population suggested by Dunlop et al. (2003).

This evidence is in itself enough to account for the major discrepancy
between the results of OWW02 and previous work. However, in this Section have
have only applied an average Doppler boosting correction factor to the
flat-spectrum sample as a whole. Obviously this average correction
factor will constitute an overestimate, or underestimate, depending
on the orientation of each individual object. 
In the next section
we proceed to consider the likely effect upon the estimated black-hole
masses of the flat-spectrum quasars due to their inclination close to
the line of sight.

\section{The Geometry of the Broad-Line Region}\label{sec:reanalysis2}

An indication of when a quasar is `misaligned' may also come
from the FWHM of the Balmer broad lines. The naive assumption is that
narrow ($\leq 4000$~km~s$^{-1}$) broad lines imply black holes of lower
mass. However, there is a wealth of evidence in the literature which
supports the view that the BLR has a disk-like geometry,
at least for radio-loud sources (e.g. Brotherton 1996). 

To account for a disk-like geometry we use a low-frequency
radio selected quasar survey to predict what the mean FWHM of the 
broad Balmer lines should be, given no spectral-index 
selection criteria. We use the quasars from the Molonglo Quasar 
sample (MQS; Kapahi et al. 1998) for which line-width measurements 
are available in the literature (Baker et al. 1999). The mean FWHM of 
the H$\beta$ line in the MQS is $\approx 7000$~km~s$^{-1}$. In
contrast, the mean FWHM of the Balmer lines in the OWW02 flat-spectrum sample 
is $\sim 3500$~km~s$^{-1}$. We therefore choose to adopt 
a correction factor of two for the flat-spectrum FWHMs to compensate 
for orientation effects. Given that $M_{\rm BH} \propto {\rm FWHM^{2}}$, this increases the black-hole mass estimates for 
the flat-spectrum sample by a factor of four.  The predicted position
of the flat-spectrum quasars on the $M_{\rm BH}-L_{\rm rad}$ plane 
after application of the inclination correction is shown 
in Fig.~\ref{fig1}, from which it can be seen that the
flat-spectrum quasars are now even more consistent with the upper and
lower radio power envelopes suggested by Dunlop et al. (2003).

\section{Conclusions}\label{sec:implications} 
We have re-analysed the data of Oshlack et al. (2002) on a sample of
flat-spectrum radio-loud quasars. Contrary to their conclusions we
find that, by correcting for the effects of inclination upon both the
radio luminosity and estimated black-hole mass, the black holes
harboured by intrinsically powerful flat-spectrum quasars are of
comparable mass to those found in other quasars of similar {\it
intrinsic} radio luminosity, i.e. $M_{\rm BH} > 10^{8}$~M$_{\circ}$.
We also find that although many of the flat-spectrum quasars
occupy the region of intrinsic radio luminosity comparable to the FRII
radio sources found in low-frequency selected
radio surveys, some of the sources may occupy the
lower-luminosity regime of radio-intermediate and radio-quiet quasars.
Therefore, we conclude that by consideration of source inclination and
intrinsic radio power, flat-spectrum quasars may well be consistent with the $L_{\rm rad} \propto
M_{\rm BH}^{2.5}$ relation found in previous studies.

\end{document}